\documentclass[epj]{webofc}
\usepackage[varg]{txfonts}   
\usepackage{hyperref}

\renewcommand*{\eqref}[1]{Eq.~(\ref{eq:#1})}

\newcommand*{\figref}[1]{Fig.~(\ref{fig:#1})}
\newcommand*{\figlab}[1]{\label{fig:#1}}

\newcommand*{\secref}[1]{Section~\ref{sec:#1}}
\newcommand*{\seclab}[1]{\label{sec:#1}}
\wocname{ARENA-2016}
\woctitle{ARENA-2016}
\begin{document}
%

\title{Overview of lunar detection of ultra-high energy particles and new plans for the SKA}
%
%

\author{\firstname{Clancy W.} \lastname{James}\inst{1}\fnsep\thanks{\email{clancy.james@physik.uni-erlangen.de}}
\and \firstname{Jaime} \lastname{Alvarez-Mu\~niz}\inst{2}
\and \firstname{Justin D.} \lastname{Bray}\inst{3}
\and \firstname{Stijn} \lastname{Buitink}\inst{4}
\and \firstname{Rustam D.} \lastname{Dagkesamanskii}\inst{5}
\and \firstname{Ronald D.} \lastname{Ekers}\inst{6} 
\and \firstname{Heino} \lastname{Falcke}\inst{7,8}
\and \firstname{Ken} \lastname{Gayley}\inst{9}
\and \firstname{Tim} \lastname{Huege}\inst{10}
\and \firstname{Maaijke} \lastname{Mevius}\inst{11}
\and \firstname{Rob} \lastname{Mutel}\inst{9}
\and \firstname{Olaf} \lastname{Scholten}\inst{11,12}
\and \firstname{Ralph} \lastname{Spencer}\inst{3}
\and \firstname{Sander} \lastname{ter~Veen}\inst{7}
\and \firstname{Tobias} \lastname{Winchen}\inst{4}
}

\institute{ECAP, Univ.\ of Erlangen-Nuremberg, 91058 Erlangen, Germany
\and
Depto.\ de F\'isica de Part\'iculas \& Instituto Galego de F\'isica de Altas Enerx\'ias, Univ.\ de Santiago de Compostela, 15782 Santiago de Compostela, Spain
\and
School of Physics \& Astronomy, Univ.\ of Manchester, Manchester M13 9PL, United Kingdom
\and
Astrophysical Institute, Vrije Universiteit Brussel, Pleinlaan 2, 1050 Brussels, Belgium
\and
LPI, Russian Academy of Sciences, Moscow Region 142290, Russia
\and
CSIRO Astronomy \& Space Science, NSW 2122, Australia
\and
Dept.\ of Astrophysics/IMAPP, Radboud Univ.\ Nijmegen, 6500 GL Nijmegen, The Netherlands
\and
Netherlands Institute for Radio Astronomy (ASTRON), 7990 AA Dwingeloo, The Netherlands
\and
Dept.\ of Physics \& Astronomy, Univ.\ of Iowa, IA 52242, USA
\and
IKP, Karlsruhe Institut f\"ur Technologie, Postfach 3640, 76021 Karlsruhe, Germany
\and
Kernfysisch Versneller Instituut,, Univ.\ of Groningen, 9747 AA Groningen, The Netherlands
\and
Interuniversity Institute for High-Energy, Vrije Universiteit Brussel, Pleinlaan 2, 1050 Brussels, Belgium
}

\abstract{%
 The lunar technique is a method for maximising the collection area for ultra-high-energy (UHE) cosmic ray and neutrino searches. The method uses either ground-based radio telescopes or lunar orbiters to search for Askaryan emission from particles cascading near the lunar surface. While experiments using the technique have made important advances in the detection of nanosecond-scale pulses, only at the very highest energies has the lunar technique achieved competitive limits. This is expected to change with the advent of the Square Kilometre Array (SKA), the low-frequency component of which (SKA-low) is predicted to be able to detect an unprecedented number of UHE cosmic rays.

In this contribution, the status of lunar particle detection is reviewed, with particular attention paid to outstanding theoretical questions, and the technical challenges of using a giant radio array to search for nanosecond pulses. The activities of SKA's High Energy Cosmic Particles Focus Group are described, as is a roadmap by which this group plans to incorporate this detection mode into SKA-low observations. Estimates for the sensitivity of SKA-low phases 1 and 2 to UHE particles are given, along with the achievable science goals with each stage. Prospects for near-future observations with other instruments are also described.
}
\maketitle
\section{Introduction}
\seclab{intro}

The lunar Askaryan technique began with Askaryan's predictions for coherent radio--microwave radiation arising from a build-up of a negative charge excess in particle cascades, and his identification of the lunar regolith as a suitable target material due to its relative radio transparency \cite{askaryan1962}. Its modern experimental incarnation was first proposed by \cite{dagkesamanskii1989}, who noted that by using ground-based radio-telescopes, the total effective area of the visible lunar surface --- approximately $20$ million km$^2$ --- could be utilised as a particle detector. The large observation distance (mean of $3.8 \,10^{5}$\,km) however imposes a high detection threshold, meaning the technique is ideally suited to targeting the rarest ultra-high-energy (UHE; $\gtrsim 10^{18}$\,eV) cosmic rays (CR) and neutrinos. Observations of UHE CR are motivated by the as-yet unknown origin of these extreme particles, while UHE neutrinos are predicted from specific theories of topological defects and super-heavy dark matter \cite{GU,KachRev,SHDM1,SHDM2}. It should be noted that efforts to observe the lunar regolith with a lunar orbiter are ongoing \cite{LORD} --- however, both the bulk of experimental efforts, and these proceedings, concentrate on the ground-based method.

Since the first search for lunar Askaryan pulses at Parkes in 1995 \cite{Parkes95}, several experiments have been performed, at Goldstone (`GLUE'; \cite{GLUE}), Kalyazin (`RAMHAND'; \cite{beresnyak2005}), the VLA (`RESUN'; \cite{jaeger2010}); Lovell (`LaLUNA'; \cite{spencer2010}); Westerbork (`NuMoon'; \cite{buitink2010}); and with ATCA and Parkes (`LUNASKA'; \cite{LUNASKA_atca,LUNASKA_parkes,LUNAKSA_Parkes_limit}).

While these experiments have produced a number of limits on high-energy particle fluxes, only the diffuse flux limits from NuMoon \cite{buitink2010}, and LUNASKA limits on Centaurus A \cite{LUNAKSA_Parkes_limit,LUNASKA_cena}, are competitive against other experiments, and then only at extremely high energies ($\gtrsim 10^{22}$~eV). This is due to two factors: no observation has yet exceeded $200$~hr, whereas most competing particle-detection experiments operate continuously; and because detection thresholds have mostly been near $10^{21}$\,eV. 


The current experimental emphasis is therefore on performing long-duration observations to limit the UHE neutrino flux, and on increasing sensitivity to detect the known flux of UHE CR. Unlike neutrinos, most of which will penetrate too deeply into the Moon to be detected, and deposit only $\sim 20$\% of their energy in detectable hadronic cascades, cosmic rays will interact immediately at the surface with $100$\% energy deposition. While it was long-thought that formation zone effects would suppress the Askaryan emission from near-surface cascades and make cosmic rays undetectable with the technique, the rapid rise and fall of the excess charge will still result in radiation regardless of the presence of a dielectric \cite{endpoints, numoon_cr}. Furthermore, future experiments --- in particular, with the Square Kilometre Array --- promise the order-of-magnitude increase in experimental sensitivity required to detect the known flux of cosmic rays, potentially allowing for arrival-direction studies.

In this overview, the main emphasis is on the planned observation strategy and simulated sensitivity of the Square Kilometre Array to UHE particles. The specific plans of the SKA's High Energy Cosmic Particles Focus Group (HECP FG\footnote{http://astronomers.skatelescope.org/home/focus-groups/high-energy-cosmic-particles/}) for observations with the low-frequency component of SKA Phase 1 will be described --- prospects for future observations with both SKA Phase 2 and other instruments will also be discussed.

\section{The Square Kilometre Array}
\seclab{sec:ska}

The Square Kilometre Array (SKA) is a radio telescope project consisting of three separate arrays covering different frequency ranges.\footnote{www.skatelescope.org} SKA-low ($50$--$350$\,MHz) will be built in Australia, while SKA-mid dishes ($\gtrsim 1$\,GHz) and an aperture array (to cover the intervening frequencies) will be built in South Africa and nearby countries. Construction of Phase $1$ of the SKA will begin shortly, with a target completion date of 2023; the full instrument in Phase 2 is targeted for completion in 2030. Phase 1 of SKA-low (SKA1-low) will consist of $131,072$ log-periodic dipole antennas deployed in $512$ stations of $256$ antennas each \cite{SKA_baseline_design_v2}, and will be deployed with a large central concentration of detectors over a $\sim$1\,km$^2$ area.

The science goals of the SKA are diverse, and are broadly categorised into Science Working Groups. The SKA is designed primarily as an imaging telescope, but the ability to observe in other modes --- pulsar searches and timing, and transient studies --- is also part of standard design features. In order to enable the SKA to perform lunar Askaryan observations, members of most prior experimental collaborations have formed the SKA's HECP Focus Group, together with groups interested in performing observations of extensive air showers \cite{ska_eas}. This group has studied the sensitivity of the SKA when using the lunar technique, with the main science case being published in the latest edition of the SKA Science book \cite{ska_lunar}. As a result of these studies, which are described further in \secref{sec:simulations}, the group has decided to focus on SKA-low. More importantly, the group has determined the required engineering changes in order to allow Phase 1 of SKA low (`SKA1-low') to perform lunar Askaryan observations, and the target observation mode is presented in \secref{sec:phase1}.

\section{Simulating the sensitivity of the SKA}
\seclab{sec:simulations}

The first detailed estimates of the sensitivity of the SKA to UHE particles were made by \cite{JamesProtheroe09a}, and have undergone continuous updates as the design of the instrument became more concrete. The current expectation is a sensitivity of $500$ m$^2$K$^{-1}$ for SKA1-low and perhaps $4000$ m$^2$K$^{-1}$ for Phase 2. This sensitivity is applicable to the detection of Askaryan pulses only when all telescope elements are correctly phased (added coherently in the voltage domain) in the direction of the pulse, and the signal power is contained to only one polarisation channel.\footnote{Equivalently, the frequency-dependence of the polarisation can be predicted, such that a signal search competes against the effective noise of one polarisation channel only.}

This `raw' sensitivity however tends to be limited by the sensitivity that can be brought to bear for online triggering, the details of which depend on the exact experimental setup. Such effects include self-vetoing due to strong signals exceeding noise-veto criteria, imperfect coverage of the lunar disk by the trigger beams, approximate dedispersion, artefacts of (analogue or digital) signal-processing, and non-optimal search algorithms  --- see \cite{singh2012, LUNASKA_atca, Bray_revised_2016, Bray_these} for further details. Only the self-veto and lunar coverage effects are reproduced here.

In the case of SKA1-low, $16$ coherent beams will be available for real-time searches (see \secref{sec:phase1}). The best trade-off between beam width and lunar coverage is expected by beamforming with all core stations (approximately $50$\% of the total), here modelled using a 2D Gaussian distribution of antenna density with 1/e fall-off radius $240$\,m.
Since the sky temperature rises rapidly at low frequencies for all pointing positions, observations in a $250$\,MHz band above $100$\,MHz are considered, although SKA1-low will be capable of observing over all $50$--$350$\,MHz simultaneously.

A preliminary optimisation of the beam pointing positions for the Moon at zenith, and assuming a frequency-independent sky plus system noise core sensitivity of $250$\,m$^2$\,K$^{-1}$, indicates that beams placed $0.9^{\prime}$ inside the limb would be optimal (see \figref{fig:aeff}, left). The event rate is evaluated assuming independently polarised triggers on each beam, for a total of $32$ independent triggers.

\begin{figure}[h]
\centering
\includegraphics[width=0.49\textwidth]{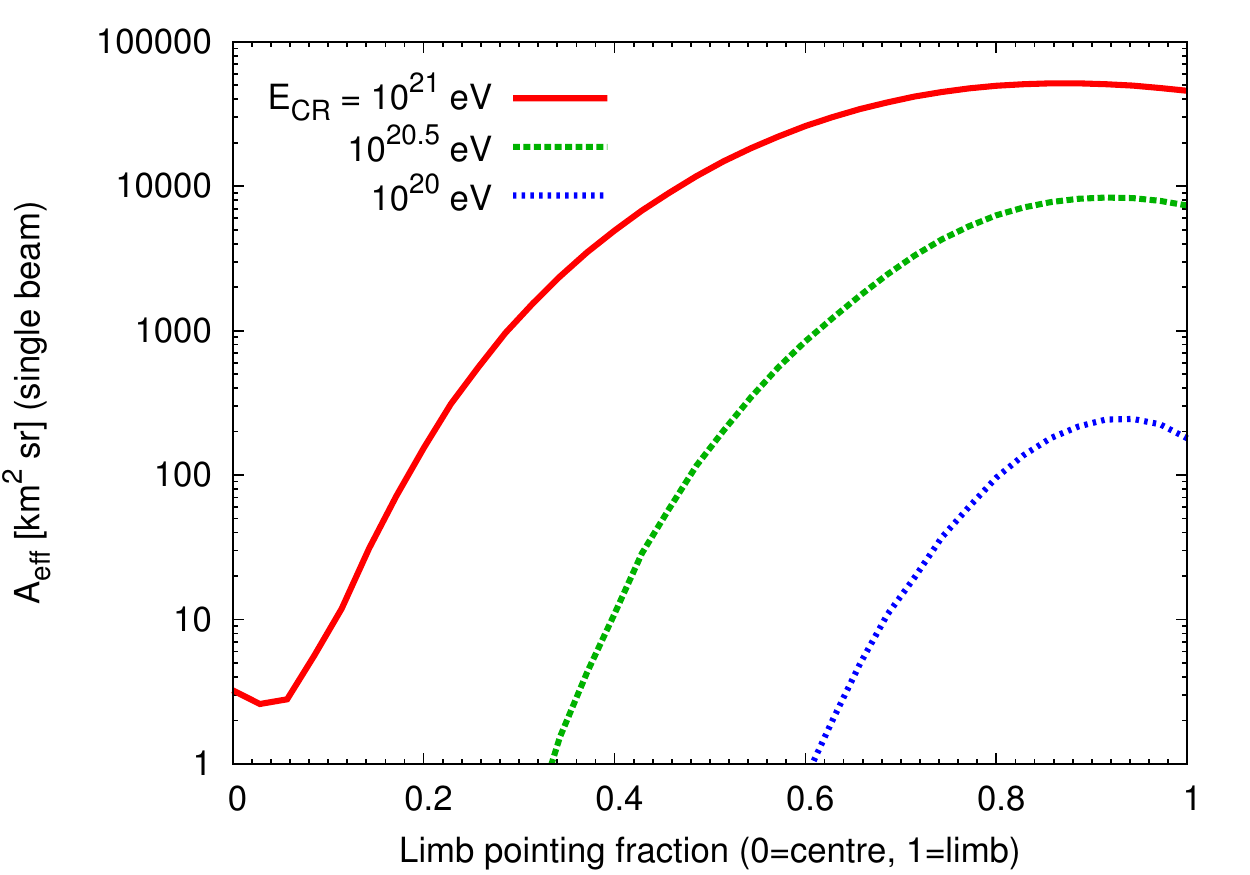} \includegraphics[width=0.49\textwidth]{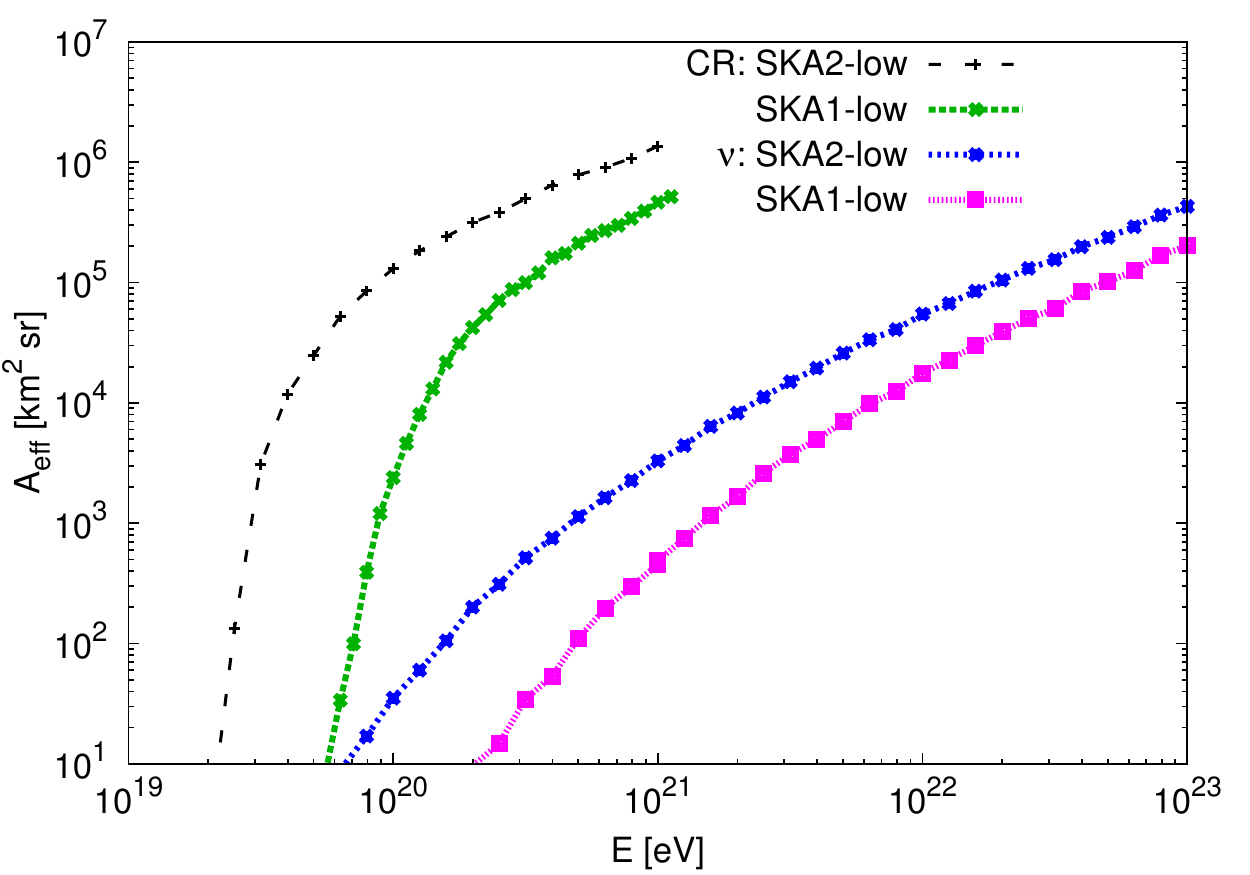}
\caption{(left) Effective area for a single SKA1-low beam (see text) as a function of fractional lunar pointing radius, for three different cosmic ray energies $E_{\rm CR}$. (right) Effective aperture $A_{\rm eff}$ of all beams of SKA2-low and SKA1-low to cosmic rays and neutrinos as a function of their energy. \cite{ska_lunar}.
} \figlab{fig:aeff}
\end{figure}

In the case of Phase 2, $100$\% lunar coverage at the sensitivity of the full array ($4000$\,m$^2$K$^{-1}$) is assumed. The chosen threshold of $10 \sigma$ corresponds to a final false positive rate of less than once per century.

\section{Results for cosmic rays and neutrinos}

Using these values, the resulting estimates of SKA1-low and SKA2-low sensitivity to UHE CR and neutrinos are shown in \figref{fig:aeff} (right). While SKA1-low will be sensitive to cosmic rays below $10^{20}$\,eV, the effective aperture in this range is significantly less than what is obtained by the Pierre Auger Observatory \cite{Abraham2010}. Using an approximate parameterisation of the cosmic ray spectrum, and a lunar visibility of $29\%$ above an elevation angle of $30^{\circ}$ from the SKA-low site, it is estimated that SKA1-low could detect of order $1$ CR yr$^{-1}$ of observing whenever the Moon is visible ($\sim$2500\,hr) \cite{james_icrc_2015}, which may allow for a first detection. The estimated rate for Phase 2 however will be much higher: of order $100$\,yr$^{-1}$, of which half will be above the 56\,EeV threshold found by the Pierre Auger Observatory to be most significant for anisotropy studies \cite{PAO07C1,Auger_cena_2015}. The sensitivity of SKA-low in either phase to UHE $\nu$ however will not be sufficient to detect the estimated flux of cosmogenic neutrinos resulting from cosmic ray interactions with background photon fields.

The clear goal therefore of lunar Askaryan observations with SKA Phase 2 will be to perform directional studies of cosmic rays at the highest energies. Preliminary estimates indicate an angular resolution of order $5^{\circ}$ will be achievable \cite{james_icrc_2015}, which is sufficient to study objects such as Centaurus A. No resolution on primary composition however is anticipated, and energy resolution may be poor, so SKA observations should not be seen as a replacement for surface arrays such as the Pierre Auger Observatory or Telescope Array.

Observations with SKA1-low will, however, be able to probe beyond-the-standard-model physics at energies above $10^{20}$\,eV. These observations could also be performed through very long ($\gg 1,000$\,hr) observations with less-sensitive (and hence less-subscribed) instruments such as AuScope at $2.3$\,GHz \cite{Bray_revised_2016}, or LOFAR observations in UHEP mode \cite{Winchen}. A first detection of cosmic rays may also be possible with instruments such as FAST \cite{FAST} or Parkes \cite{Bray_revised_2016}.

\section{Implementation with Phase 1 of SKA-low}
\seclab{sec:phase1}

\begin{figure}[h]
\centering
\includegraphics[width=0.6\textwidth]{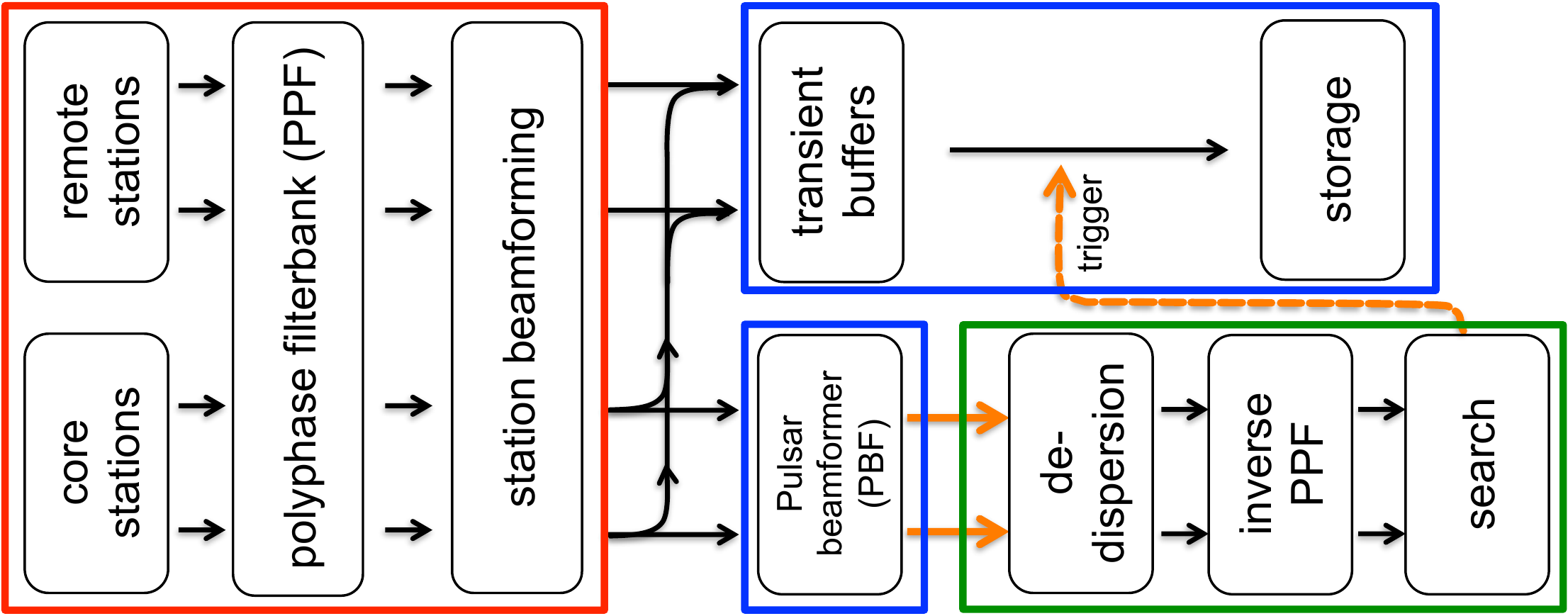}
\caption{Sketch of the proposed signal path for SKA1-low lunar Askaryan observations, colour-coded by observation mode. Red: part of the standard imaging pipeline; blue: added functionality for pulsar and transient observations; green: specialised lunar pulse detection hardware; orange: required interfaces.}
\figlab{fig:signal_path}
\end{figure}

A sketch of the proposed SKA-lunar signal path for SKA1-low is shown in \figref{fig:signal_path}, based on the updated SKA1-low baseline design \cite{SKA_baseline_design_v2}. In standard operations, dual linear polarisations from all 256 elements in each station will be digitised, passed through a polyphase filter (PPF) to split them into $300$ $1$\,MHz channels, and formed into a `station beam'. These station beams can then be passed to the correlator for imaging (not shown), the pulsar timing array beamformer (PBF) to form `array' beams, and/or buffered to memory for transient detection. The PBF will be able to produce 16 dual polarisation array beams at $300$\,MHz bandwidth.. In order to search for lunar Askaryan pulses, these beams will be directed at the Moon, and their output passed to dedicated hardware.

To correct for ionospheric delays, each sub-band will undergo a bulk de-dispersion according to the best real-time STEC prediction, and then be passed to an inverse PPF. This will operate on each station beam to transform the signal back to the time-domain. Note that the bulk dedispersion, and also any noise-whitening, can be performed for free by modifying the initial inverse PPF coefficients.

In order to search for a lunar Askaryan signal, each beam will be copied to many separate streams, with each applying a different fine dedispersion to account for errors in the initial bulk dedispersion. Each of the subsequent signals will then be searched for possible lunar pulses.

The optimal search method for a coherent signal against white noise uses a matched filter. However, even if the pulse shape can be perfectly predicted, the signal phase will vary with respect to the sampling phase. Therefore, a search over multiple signal shapes must be performed. The expected trigger rate from pure noise events is expected to be in the range 0.1--1\,Hz.

The final stage will be an anti-coincidence test over all $16$ beams to veto RFI events. Once a trigger is generated that passes the veto, both the array beams, and the buffered station beams, will be stored offline for further analysis. Further details of the proposed implementation of these components is given in \cite{lunar_technical}.

The doubled effective area available in offline processing through buffered data from non-core stations will be sufficient to discriminate these false positives from true lunar events --- which necessarily implies that the final sensitivity will be trigger-limited. Returning $10$\,$\mu$s per trigger is expected to be sufficient to capture a fully dispersed pulse, account for arrival-time uncertainties on long baselines, and provide sufficient off-signal data to estimate the noise level. The total data volume, assuming station-beam data at 8-bit resolution, will be $8$\,MB per trigger.

\section{Conclusion}

Using the lunar Askaryan technique, the low-frequency component of the Square Kilometre Array in Phase 2 will have the ability to detect an unprecedented number of ultra-high-energy cosmic rays, in order to perform directional studies and potentially resolve the cosmic ray origin. Almost all the necessary technical development has been performed by the member groups of the High Energy Cosmic Particles Focus Group within the SKA Organisation. Observations with LOFAR's UHEP mode, and with Phase 1 of SKA-low, will provide both critical technical development, and important limits on beyond-the-standard-model physics. There are good prospects for a first detection with SKA1-low or other instruments, and for very-long-duration observations specifically to test exotic models of extremely high-energy particles.




%
%
%

\end{document}